\documentclass[english,utf8x]{article-hermes}

\usepackage{xspace}
\usepackage{arabtex}
\usepackage{array}
\usepackage{tikz}
\usepackage{url}
\usepackage{booktabs}
\usepackage{multirow}
\usepackage{mathabx}
\usepackage{utf8}
\usepackage{verbatim}
\usepackage{fancyvrb}
\usepackage{relsize}
\usepackage{wasysym}
\def\framework{\textsc{MERF}\xspace}
\newcolumntype{L}[1]{>{\raggedright\let\newline\\\arraybackslash\hspace{0pt}}m{#1}}
\newcolumntype{C}[1]{>{\centering\let\newline\\\arraybackslash\hspace{0pt}}m{#1}}
\newcolumntype{R}[1]{>{\raggedleft\let\newline\\\arraybackslash\hspace{0pt}}m{#1}}

\newcommand{\cci}[1]{{\small \texttt{#1}}}
\newcommand{\utfrl}[1]{\setcode{utf8}\RL{#1}\setcode{standard}}
\newcommand{\notrutfrl}[1]{\transfalse\setcode{utf8}\RL{#1}\setcode{standard}\transtrue}
\newcommand{\notrrl}[1]{\transfalse\RL{#1}\transtrue}
\newcommand{\noarrl}[1]{\arabfalse\RL{#1}\arabtrue}

\journal{TAL. Volume 58 -- n° 3/2017}{97}{121}

\title[\framework]{Morphology-based Entity and Relational Entity Extraction Framework for Arabic}

\vspace*{-.5em}
\author{Amin Jaber\fup{*} \andauthor Fadi A. Zaraket\fup{**}}

\address{
\fup{*} Purdue University, West Lafayette, IN\\
\fup{**} American University of Beirut, Beirut 1107 2020, Lebanon}

\abstract{
Rule-based techniques to extract relational entities from 
documents allow users to specify desired entities with 
natural language questions, finite state automata, regular expressions and
structured query language. 
They require linguistic and programming expertise and 
lack support for Arabic morphological analysis.
We present a morphology-based entity and relational entity extraction 
framework for Arabic (\framework).
\framework requires
basic knowledge of linguistic features and regular expressions, 
and provides the ability to interactively specify
Arabic morphological and synonymity features, 
tag types associated with regular expressions, and
relations and code actions defined over matches of subexpressions. 
\framework constructs entities and relational entities from matches 
of the specifications.
We evaluated \framework with several case studies.
The results show that \framework requires shorter development time and effort compared to existing application specific techniques and produces 
reasonably accurate results within a reasonable overhead in run time.
}


\motscles{
Arabe,
extraction d'information,
traitement du langage naturel,
marquage.
\vspace*{-.5em}
}

\resume{
Les techniques à base de règles pour extraire des entités permettent 
de spécifier les entités souhaitées en utilisant des questions 
de langage naturel, des automates à états finis, des expressions régulières et des 
instructions d'extraction de données.
Ils nécessitent des expertises en linguistique et en programmation, et
ne soutiennent pas l'analyse morphologique de l'arabe.
On présente pour l'arabe 
un cadre d'extraction d'entité 
renforcé par l'analyse morphologique (\framework). 
Il exige des connaissances de base des 
caractéristiques linguistiques et des expressions régulières, 
et fournit la possibilité de spécifier de façon interactive 
des fonctionnalités de morphologie et synonymie arabes,
des types de tag associés avec des expressions régulières, 
et des relations et actions de code définies sur les correspondances de sous-expressions.
\framework construit des entités relationnelles à partir des correspondances
des spécifications.
On évalue \framework avec des études de cas.
Les résultats montrent que \framework nécessite un effort de développement plus court par rapport aux techniques existantes 
et produit des résultats raisonnablement précis avec une surcharge raisonnable en temps d'exécution.}

\begin{document}

\maketitlepage

\section{Introduction}
\label{sec:introduction}

{\em Computational Linguistics} (CL) is concerned with building 
accurate linguistic computational models.
{\em Natural Language Processing} (NLP) is concerned with automating the 
understanding of natural language. 
CL and NLP tasks range from simple ones such as spell checking and typing error correction to more complex tasks including 
{\em named entity recognition} (NER), {\em cross-document analysis}, 
{\em machine translation}, and {\em relational entity extraction}~\cite{linckels2011natural,ferilli2011natural}.
Entities are elements of text that are of interest to an NLP task.
Relational entities are elements that connect entities.
{\em Annotations} relate chunks of text to {\em labels} denoting 
semantic values such as entities or relational entities.
We refer to annotations and labels as {\em tags} and {\em tag types}, 
respectively, in the sequel.

Supervised and unsupervised empirical learning techniques tackle NLP and CL tasks.
They employ machine learning without the need to manually encode the requisite knowledge~\cite{soudi2007arabic}. 
Supervised learning techniques require training corpora annotated 
with {\em correct} tags to learn a computational model.
Supervised and unsupervised techniques require annotated reference corpora to 
evaluate the accuracy of the technique using metrics such as precision and 
recall~\cite{englishtreebank,arabictreebank,chinesetreebank}. 

Researchers build 
training and reference corpora 
either manually, incrementally using learning techniques, or using knowledge-based 
annotation techniques that recognize and extract entities and relational 
entities from text. 
Knowledge-based techniques use linguistic and 
rhetorical domain specific knowledge encoded into sets of rules 
to extract entities and relational entities~\cite{soudi2007arabic}.
While existing annotation, entity, and relational entity 
extraction tools exist
\cite{chiticariu2010systemt
,atzmueller2008rule
,urbain2012user%
,settles2011closing
,mmax2
,brat
},
most of them lack Arabic language support, and almost all of them
lack Arabic morphological analysis support
~\cite{habash2006arabic}. 
Fassieh~\cite{attia2009fassieh} is a {\em commercial} Arabic annotation tool
with morphological analysis support and text factorization. 
However, this tool lacks support for entity and relational entity extraction.

\transtrue
\setcode{utf8}
\setarab
\transfalse
\begin{figure}[tb]
\begin{center}
\relsize{-0.5}
\resizebox{\columnwidth}{!}{
    \begin{tabular}{|C{10cm}C{8cm}|} 
      \hline
  \begin{tabular}{R{10cm}}
\RL{تقود} \RL{وأنت} $\stackrel{u1}{\framebox[1.2\width]{\RL{الأول}}}$ $\stackrel{p_2}{\framebox[1.2\width]{\RL{التقاطع}}}$ \RL{من} $\stackrel{r_1}{\framebox[1.2\width]{\RL{بالقرب}}}$ $\stackrel{n_1}{\framebox[1.2\width]{\RL{خليفة}}}$ $\stackrel{p_1}{\framebox[1.2\width]{\RL{برج}}}$ \RL{تلاحظ} \RL{ألا} \RL{المستحيل} \RL{من} 
  \\ ~ \\
$\stackrel{r_3}{\framebox[1.2\width]{\RL{فيها}}}$ \RL{تسلك} \RL{التي} $\stackrel{u_2}{\framebox[1.2\width]{\RL{الأولى}}}$ \RL{المرّة} \RL{هذه} \RL{كانت} \RL{وإن} \RL{حتّى} $\stackrel{n_3}{\framebox[1.2\width]{\RL{زايد،}}}$ $\stackrel{n_2}{\framebox[1.2\width]{\RL{الشيخ}}}$ $\stackrel{p_3}{\framebox[1.2\width]{\RL{شارع}}}$ $\stackrel{r_2}{\framebox[1.2\width]{\RL{في}}}$ \RL{سيارتك}
  \\ ~ \\
\RL{العالم.} \RL{في} \RL{الأطول} \RL{يُعد} \RL{الذي}  $\stackrel{p_7}{\framebox[1.2\width]{\RL{المبنى}}}$ \RL{هذا} \RL{من} $\stackrel{r_4}{\framebox[1.2\width]{\RL{مقربة}}}$ \RL{على} $\stackrel{p_6}{\framebox[1.2\width]{\RL{مول}}}$ $\stackrel{p_5}{\framebox[1.2\width]{\RL{دبيّ}}}$ \RL{يقع} $\stackrel{p_4}{\framebox[1.2\width]{\RL{الطريق؛}}}$ \RL{هذا}
  \\ ~ \\
  \multicolumn{1}{L{10cm}}{
   \arabfalse \transtrue
   \RL{من المستحيل ألا تلاحظ برج خليفة بالقرب من التقاطع الأول وأنت تقود سيارتك في شارع الشيخ زايد، حتّى وإن كانت هذه المرّة الأولى التي تسلك فيها هذا الطريق؛ يقع دبيّ مول على مقربة من هذا المبنى الذي يُعد الأطول في العالم.}
   \arabtrue \transfalse} 
  \\ ~ \\
  \multicolumn{1}{L{10cm}}{It is impossible not to notice $\stackrel{n1}{\framebox[1.2\width]{Khalifa}}$ $\stackrel{p_1}{\framebox[1.2\width]{Tower}}$ $\stackrel{r_1}{\framebox[1.2\width]{next to}}$ the $\stackrel{u_1}{\framebox[1.2\width]{first}}$ $\stackrel{p_2}{\framebox[1.2\width]{intersection}}$ while you are driving your car $\stackrel{r_2}{\framebox[1.2\width]{on}}$ $\stackrel{n_2}{\framebox[1.2\width]{Sheikh}}$ $\stackrel{n_3}{\framebox[1.2\width]{Zayed}}$ $\stackrel{p_3}{\framebox[1.2\width]{Road}}$, even if this was $\stackrel{u_2}{\framebox[1.2\width]{the first}}$ time that you take this $\stackrel{p_4}{\framebox[1.2\width]{road}}$; $\stackrel{p_5}{\framebox[1.2\width]{Dubai}}$ $\stackrel{p_6}{\framebox[1.2\width]{Mall}}$ is located $\stackrel{r_4}{\framebox[1.2\width]{near}}$ this $\stackrel{p_7}{\framebox[1.2\width]{building}}$, which is the longest in the world.} 
  \end{tabular}
&
\resizebox{0.6\columnwidth}{!}{
  \input{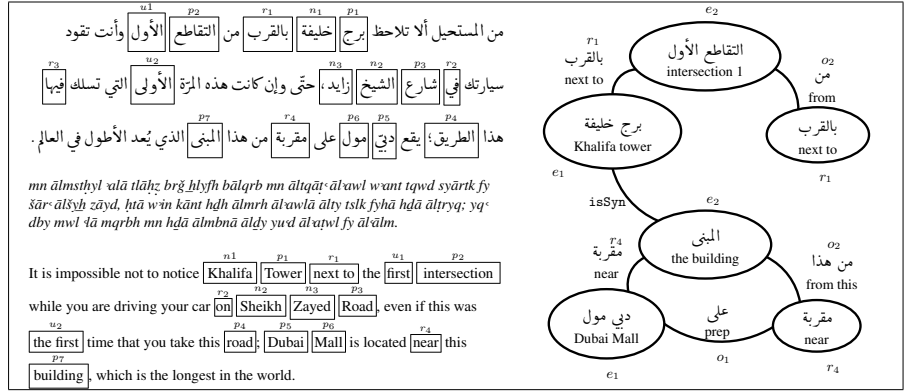}}
\\  \hline
    \end{tabular}
}
\end{center}
\vspace{-1em} 
  \caption{Direction example with Arabic text, annotated with entities, transliteration, translation, and extracted relational entities in a graph.} 
  \label{fig:intromotiv}
\vspace{-1em} 
\end{figure}
\transtrue
\setcode{standard}

Figure~\ref{fig:intromotiv} illustrates the target of \framework using 
the directions to Dubai Mall example~\footnote{Text taken from the Dubai 
Mall website ~\url{http://www.thedubaimall.com/ar/}.}.
The figure also presents a transliteration and an English translation of 
the Arabic text.
The framed words in the text are entities referring to names of people 
($n_1,n_2,n_3$), 
names of places ($p_1,\dots,p_7$), 
relative positions ($r_1,\dots,r_4$), 
and numerical terms ($u_1,u_2$). 
We would like to extract those entities, and then extract
the relational entities forming the graph in Figure~\ref{fig:intromotiv} 
where vertices express entities, 
and edges represent the relational entities.


In this paper, 
we present \framework, a morphology-based entity and relational entity
extraction framework for Arabic text.
\framework provides a user-friendly interface where the user defines tag types 
and associates them with
regular expressions over 
Boolean formulae.
A Boolean formula is defined by a term, negation of a term, or disjunction of terms.
Terms are matches to Arabic morphological features including 
prefix, stem, suffix, part of speech (POS) tags, gloss tags, extended synonym 
tags, and semantic categories.
For example, entity $p_1$ in Figure~\ref{fig:intromotiv} has a ``place'' semantic 
category.
\framework regular expressions support operators such as concatenation, 
zero or one, zero or more, one or more, up to $M$ repetitions where $M$ is 
a non-zero positive integer, and logical conjunction and disjunction.
For example, the sequence between $p_1$ and $p_2$ matches a regular expression $re$ 
that requires two semantic place categories with a place-preposition POS 
tag ($r_1$) in between.

An editor allows the user to associate an action with each subexpression.
The user specifies the action with C++ code and uses an API to access information 
related to the matches such as text, position, length, morphological features, 
and numerical value.
Each regular expression is associated with a named identifier to form a 
{\em local grammar} like structure~\cite{traboulsi2009arabic}. 
A relation definition GUI allows the user to provide relational tuples 
where each tuple has a 
source, a destination and an edge label.
The user uses the regular expression identifiers to define the relational 
tuple elements.
For example, the relation between $e_1, e_2$ and $r$ shown in 
Figure~\ref{fig:intromotiv} is a match of a relational tuple 
over the components of $re$. 
We refer to regular expressions and Boolean formulae as expressions and formulae, respectively.
We also refer to expressions as rules
when used in a grammar context; e.g.  when used with an identifier. 

\framework takes an Arabic text and the local grammar defined by the 
Boolean formulae and the regular expressions.
\framework computes the morphological solutions of the 
input text then computes matches to the Boolean formulae therein.
\framework then generates a {\em non-deterministic finite state 
automata} (NDFSA) for each expression and simulates it with the 
sequence of Boolean formulae matches to compute the regular 
expression matches. 
\framework generates executable code for the actions associated with
the regular expressions, 
compiles, links, and executes the generated code 
as shared object libraries.
Finally, \framework constructs the semantic relations and 
cross-reference between entities.
\framework also provides visualization tools to present the matches, 
and estimate their accuracy with respect to reference tags.

This work significantly extends~\citename{JaZaMatar}~\citeyear{JaZaMatar}
that allows for manual, and morphology annotation.
\framework enables a user to incrementally create 
complex annotations for Arabic based on automatic 
extraction of morphological tags through a user-friendly interactive interface. 
\framework has the following advantages.
\begin{itemize}
  \item \framework provides a novel and intuitive visual interface to build formulae over morphological features, 
    build regular expressions over the resulting formulae, and thereafter compute automatic tags.
  \item To our knowledge, this morphology-based framework is the first for Arabic entity and relational entity extraction.
  \item \framework provides the user with the ability to rapidly create annotated Arabic text corpora with sophisticated morphology-based tags.
\end{itemize}

In \framework, we make the following contributions.
\begin{itemize}
  \item \framework enables the user to define relations in a simple manner
    and automatically detects relational entities matching the user defined relations. 
  \item \framework enables the user to associate subexpressions
    with code actions, and executes the code action 
    when a corresponding match is found.
    It also provides an API to enable access to match 
    features such as text, position, length, numerical value, and morphological features.
  \item \framework enables the user to tag words based on a novel light Arabic WordNet relation that leverages the synonym $Syn^k$ feature.
  \item \framework is open source and available online for the research community 
    under \url{https://github.com/codelogicanalysis/atmine}.
\end{itemize}

The rest of the paper is structured as follows.
Section~\ref{sec:morph} introduces Arabic morphological analysis 
and its important role in Arabic NLP.
Section~\ref{sec:methodology} explains the methodology of \framework.
Section~\ref{sec:framework} presents \framework components.
Section~\ref{sec:gui} presents \framework GUI. 
Section~\ref{sec:related} presents and discusses related work.
Section~\ref{sec:results} presents the evaluation results. 
Finally, we conclude and discuss future work in Section~\ref{sec:conclusion}.

\section{Background: Morphological Analyzer}
\label{sec:morph}

\vocalize

\def\pp{\ensuremath{{\cal P}}} 
\def\ss{\ensuremath{{\cal S}}} 
\def\xx{\ensuremath{{\cal X}}} 
\def\PP{\ensuremath{\mathit{POS}}} 
\def\GG{\ensuremath{\mathit{GLOSS}}} 
\def\AC{\ensuremath{\mathit{CAT}}} 

Morphological analysis is key to Arabic 
NLP
due to the exceptional degree of ambiguity in writing, 
the rich morphology, and the complex word derivation 
system~\cite{arabicmorph,shahrour2016camelparser,pasha2014madamira}.
Short vowels, also known as diacritics, are typically omitted in Arabic text
and inferred by readers~\cite{habash2006arabic}. 
For example, the word \RL{bn}%
can be interpreted as \RL{bon} (``coffee'') with a {\em damma} diacritic on the 
letter \utfrl{بـ} or 
\vocalize \RL{bin} (``son of'') with a 
{\em kasra} diacritic on the letter \utfrl{بـ}.

Morphological analysis is required even for tokenization of Arabic text. 
The position of an Arabic letter in a word 
(beginning, middle, end, and standalone) changes
its visual form.
Some letters have non-connecting end forms which allows visual
word separation without the need of a white space separator. 
For example, the word \utfrl{ياسمين} can be interpreted as
the ``Jasmine'' flower, 
as well as \notrutfrl{يا} (the calling word) followed by
the word \notrutfrl{سمين} (obese). 
Consider the sentence 
\notrrl{AlmdrsT}\notrrl{_dhb alwald-il_A}
\arabfalse \RL{_dhb alwald-il_A almdrsT} \arabtrue
(``the kid went to school''). 
The letters \notrutfrl{د} and \notrutfrl{ى} have 
non-connecting end of word forms and the words 
\notrutfrl{الولد},\notrutfrl{الى}, and\notrutfrl{المدرسة} 
are visually separable, 
yet there is no space character in between.
Newspaper articles with text justification requirements, 
SMS messages, and automatically digitized documents
are examples where such problems occur. 

\framework is integrated with {\em Sarf}, 
an in-house open source Arabic morphological analyzer based on 
finite state transducers~\cite{ZaMaColing2012DemosSarf}. 
Given an Arabic word, Sarf returns 
a set of morphological solutions. 
A word might have more than one solution 
due to multiple possible segmentations and multiple tags associated 
with each word. 
A morphological solution is the internal structure of the word 
composed of several morphemes including 
{\em affixes} ({\em prefixes} and {\em suffixes}), and a
{\em stem}, where each morpheme is associated with tags such as 
POS, gloss, and category tags~\cite{arabicmorph,habash2010introduction}.

Prefixes attach before the stem and a word can have multiple prefixes. 
Suffixes attach after the stem and a word can have multiple suffixes. 
Infixes are inserted inside the stem to form a new stem. 
In this work we consider a set of stems that includes infix morphological changes. 
The part-of-speech tag, referred to as POS, 
assigns a morpho-syntactic tag for a morpheme. 
The gloss is a brief semantic notation of morpheme in English. 
A morpheme might have multiple glosses as it could stand for multiple meanings. 
The category is a custom tag that we assign to multiple morphemes. 
For example, we define the {\tt Name of Person} category to include proper names.

We denote by 
\ss,
\pp,
\xx,
\PP,
\GG, and 
\AC, the set of 
all stems,
prefixes,
suffixes,
POS,
gloss, 
and user defined category tags, respectively.
Let $T=\langle t_1,t_2,\ldots,t_M\rangle$ be a set of Arabic words denoting the 
text documents.
\framework uses Sarf to compute a set of morphological solutions $M(t)=\{m_1,m_2,\ldots,m_N\}$
for each word $t\in T$. 
Each morphological solution $m\in M(t)$ is a tuple of the form 
$\langle p,s,x,P,G,C\rangle \in \pp \times \ss \times \xx \times \PP \times \GG \times \AC$ 
where $p=p_1 \ldots p_{|p|}$, $x=x_1 \ldots x_{|x|}$, 
$P=P_{p_1}\ldots P_{p_{|p|}} P_s P_{x_1}\ldots P_{x_{|x|}}$,
$G=G_{p_1}\ldots G_{p_{|p|}} G_s G_{x_1}\ldots G_{x_{|x|}}$, and
$C=C_{p_1} \ldots C_{p_{|p|}} C_s C_{x_1}\ldots C_{x_{|x|}}$. 
$P_{p_i}, G_{p_i},$ and $C_{p_i},1\le i \le |p|$ are the POS, gloss and 
category tags of prefix $p_i$. 
$P_{x_j}, G_{x_j},$ and $C_{x_j},1\le j \le |x|$ are the POS, gloss and 
category tags of suffix $x_i$. 
$P_s, G_s,$ and $C_s$ are the POS, gloss and 
category tags of stem $s$. 
Intuitively, 
$p,x,P,G$ and $C$ are concatenations of prefix, suffix, POS, gloss and
category values, respectively.

\begin{table}[tb]
  \centering
  \resizebox{0.85\columnwidth}{!}{
    \begin{tabular}{|r|c|c|c|c|c|}
          & \multicolumn{3}{c|}{\textbf{Prefixes}} & \textbf{Stem} & \textbf{Suffix} \\
    \textbf{Data} & \utfrl{فَ} & \utfrl{سَ} & \utfrl{يَ} & \utfrl{أْكُل} & \utfrl{ها} \\
    \textbf{POS} & CONJ+ & FUT+ & IV3MS+ & VERB\_IMPERFECT & IVSUFF\_DO:3FS \\
    \textbf{Gloss} & and/so & will & he/it & eat/consume & it/them/her \\
    \textbf{index} & \multicolumn{3}{c|}{10} & 13 & 16 \\
    \textbf{length} & \multicolumn{3}{c|}{3} & 3 & 2 \\
    \end{tabular}%
    }
    \vspace{-1em}
  \caption{\label{tab:samplesolution}Sample solution vector for \utfrl{فَسَيَأْكُلها}.}
\end{table}%

Table~\ref{tab:samplesolution} shows the morphological analysis
of the word \notrutfrl{فَسَيَأْكُلها}. 
The word is composed of the prefix morphemes 
\utfrl{فَ}, \utfrl{سَ}, and \utfrl{يَ}, followed by the 
stem \utfrl{أْكُل}, and then followed by the suffix morpheme
\utfrl{ها}. 
Each morpheme is associated with a number of morphological features.
The {\tt CONJ},
{\tt FUT}, 
{\tt IV3MS} 
{\tt VERB\_IMPERFECT}, and 
{\tt IVSUFF\_DO:3FS} POS tags indicate
conjunction, 
future, 
third person masculine singular subject pronoun,
an imperfect verb, and 
a third person feminine singular object pronoun, respectively.
The POS and gloss notations follow the Buckwalter notation~\cite{Buckwalter:02}.

\vspace{-1em}
\section{\framework Methodology}
\vspace{-.5em}
\label{sec:methodology}

\begin{figure}[tb!]
\centering
\resizebox{\columnwidth}{!}{
	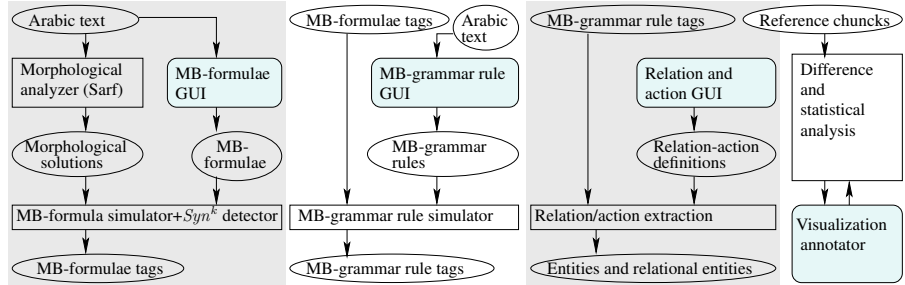
}
\vspace{-2em}
\caption{\framework four process methodology with rounded corner blocks for GUI.}
\label{f:overview}
\end{figure}

Figure~\ref{f:overview} illustrates the four processes involved in \framework methodology. 
The first process takes Arabic text and provides the user 
with a morphology-based Boolean (MB) formulae GUI. 
The user interactively composes MB-formulae using the GUI and the output 
of the simulator and the $Syn^k$ detector. 
The simulator and the detector apply the formulae over the morphological solutions 
of the Arabic text and produce the MB-formulae tags. 

The second process takes the MB-formulae tags and the Arabic text and provides
the user with a morphology-based grammar rule GUI. 
The user interactively composes MB-grammar rules using the GUI and the output 
of the MB-grammar rule simulator. 
The grammar rule simulator applies the rules over the MB-formulae tags and 
produces the MB-grammar rule tags. 

The third process takes the MB-grammar rule tags and provides the user with 
a relation and action GUI. 
The user interactively provides (1) the relation definitions and 
(2) the actions in terms of identifiers from the MB-grammar 
rules.
The relation extraction produces the target entities and relational
entities. 
The action execution enriches the entities and the relational entities
with powerful semantics.
For example, users can utilize actions to compute statistical features, 
store intermediate results, 
or apply intelligent entity inference techniques
as we show later in the numerical extraction example of 
Subsection~\ref{subsec:numnorm}.
Finally, in the fourth process the user compares the results with golden
reference chunks and visualizes the difference. 
This allows the user to refine the formulae, rules, relations and actions. 

After relation extraction, we are interested to relate 
entities that express the same concept.
\framework provides the extended synonym feature of second order as 
a default cross-reference relation ($Syn^2$).
In Figure~\ref{fig:intromotiv}, triggering this feature creates the 
edge labeled with \cci{isSyn} 
between the nodes \cci{Khalifa Tower} and \cci{The building}.

The user may refine the defined formulae, rules and relations and 
the corresponding formulae tags, rule tags, entities and relational entities
either using the GUI or directly through readable output files.
The files are in the javascript object notation (JSON)~\cite{nolan2014javascript} 
format that is intuitive to read and modify.
\framework separates the user defined formulae, rules, actions and relations in
a \framework tag type file and the 
matching tags in a tags files.
The separation serves the user to apply the tag types to multiple case studies 
and to obtain a separate file of resulting tags for each.

\vspace{-1em}
\section{\framework Components}
\label{sec:framework}
\label{sec:tool}

\subsection{The extended synonymy feature $Syn^k$}
\label{subsec:synk}

Up to our knowledge, 
$Syn^k$ provides the first light Arabic WordNet based on the lexicon of Sarf.
The sets $E, A,$ and $L$ denote all English words, Arabic words, 
and Arabic lexicon words, respectively.
Recall that $\GG$ and $\ss$ denote the set of glosses and stems in the morphological analyzer, respectively.
We have $\GG \subset E$ and $\ss \subset L \subset A$. 
Function $\alpha: \ss \rightarrow 2^{\GG}$ maps Arabic stems to 
subsets of related English glosses, 
where $2^{\GG}$ denotes the power set of $\GG$ which is the set of all subsets 
of $\GG$.
Function $\gamma: L \rightarrow 2^{\ss}$ maps Arabic lexicon words to subsets 
of relevant Arabic stems. 

\begin{figure}[tb]
\setcode{utf8}
\begin{center}
  \resizebox{0.75\columnwidth}{!}{ 
  	{\relsize{-2} 
\ifx\du\undefined
  \newlength{\du}
\fi
\setlength{\du}{15\unitlength}
\begin{tikzpicture}
\pgftransformxscale{1.000000}
\pgftransformyscale{-1.000000}
\definecolor{dialinecolor}{rgb}{0.000000, 0.000000, 0.000000}
\pgfsetstrokecolor{dialinecolor}
\definecolor{dialinecolor}{rgb}{1.000000, 1.000000, 1.000000}
\pgfsetfillcolor{dialinecolor}
\definecolor{dialinecolor}{rgb}{1.000000, 1.000000, 1.000000}
\pgfsetfillcolor{dialinecolor}
\pgfpathellipse{\pgfpoint{3.630147\du}{12.268195\du}}{\pgfpoint{0.803947\du}{0\du}}{\pgfpoint{0\du}{0.833295\du}}
\pgfusepath{fill}
\pgfsetlinewidth{0.020000\du}
\pgfsetdash{{\pgflinewidth}{0.200000\du}}{0cm}
\pgfsetdash{{\pgflinewidth}{0.200000\du}}{0cm}
\pgfsetmiterjoin
\definecolor{dialinecolor}{rgb}{0.000000, 0.000000, 0.000000}
\pgfsetstrokecolor{dialinecolor}
\pgfpathellipse{\pgfpoint{3.630147\du}{12.268195\du}}{\pgfpoint{0.803947\du}{0\du}}{\pgfpoint{0\du}{0.833295\du}}
\pgfusepath{stroke}
\definecolor{dialinecolor}{rgb}{0.000000, 0.000000, 0.000000}
\pgfsetstrokecolor{dialinecolor}
\node at (3.630147\du,12.371528\du){\RL{ماء}};
\definecolor{dialinecolor}{rgb}{1.000000, 1.000000, 1.000000}
\pgfsetfillcolor{dialinecolor}
\pgfpathellipse{\pgfpoint{7.412496\du}{12.443762\du}}{\pgfpoint{0.899916\du}{0\du}}{\pgfpoint{0\du}{0.848262\du}}
\pgfusepath{fill}
\pgfsetlinewidth{0.020000\du}
\pgfsetdash{{\pgflinewidth}{0.200000\du}}{0cm}
\pgfsetdash{{\pgflinewidth}{0.200000\du}}{0cm}
\pgfsetmiterjoin
\definecolor{dialinecolor}{rgb}{0.000000, 0.000000, 0.000000}
\pgfsetstrokecolor{dialinecolor}
\pgfpathellipse{\pgfpoint{7.412496\du}{12.443762\du}}{\pgfpoint{0.899916\du}{0\du}}{\pgfpoint{0\du}{0.848262\du}}
\pgfusepath{stroke}
\definecolor{dialinecolor}{rgb}{0.000000, 0.000000, 0.000000}
\pgfsetstrokecolor{dialinecolor}
\node at (7.412496\du,12.547096\du){\RL{نضح}};
\definecolor{dialinecolor}{rgb}{1.000000, 1.000000, 1.000000}
\pgfsetfillcolor{dialinecolor}
\pgfpathellipse{\pgfpoint{11.327704\du}{12.243392\du}}{\pgfpoint{0.887504\du}{0\du}}{\pgfpoint{0\du}{0.840661\du}}
\pgfusepath{fill}
\pgfsetlinewidth{0.020000\du}
\pgfsetdash{{\pgflinewidth}{0.200000\du}}{0cm}
\pgfsetdash{{\pgflinewidth}{0.200000\du}}{0cm}
\pgfsetmiterjoin
\definecolor{dialinecolor}{rgb}{0.000000, 0.000000, 0.000000}
\pgfsetstrokecolor{dialinecolor}
\pgfpathellipse{\pgfpoint{11.327704\du}{12.243392\du}}{\pgfpoint{0.887504\du}{0\du}}{\pgfpoint{0\du}{0.840661\du}}
\pgfusepath{stroke}
\definecolor{dialinecolor}{rgb}{0.000000, 0.000000, 0.000000}
\pgfsetstrokecolor{dialinecolor}
\node at (11.327704\du,12.346725\du){\RL{رشّ}};
\pgfsetlinewidth{0.020000\du}
\pgfsetdash{}{0pt}
\pgfsetdash{}{0pt}
\pgfsetmiterjoin
\definecolor{dialinecolor}{rgb}{0.000000, 0.000000, 0.000000}
\pgfsetstrokecolor{dialinecolor}
\pgfpathellipse{\pgfpoint{4.446635\du}{12.287500\du}}{\pgfpoint{1.753365\du}{0\du}}{\pgfpoint{0\du}{1.150000\du}}
\pgfusepath{stroke}
\definecolor{dialinecolor}{rgb}{0.000000, 0.000000, 0.000000}
\pgfsetstrokecolor{dialinecolor}
\node at (4.446635\du,12.390833\du){};
\pgfsetlinewidth{0.020000\du}
\pgfsetdash{}{0pt}
\pgfsetdash{}{0pt}
\pgfsetmiterjoin
\definecolor{dialinecolor}{rgb}{0.000000, 0.000000, 0.000000}
\pgfsetstrokecolor{dialinecolor}
\pgfpathellipse{\pgfpoint{7.512500\du}{12.275000\du}}{\pgfpoint{2.487500\du}{0\du}}{\pgfpoint{0\du}{1.237500\du}}
\pgfusepath{stroke}
\definecolor{dialinecolor}{rgb}{0.000000, 0.000000, 0.000000}
\pgfsetstrokecolor{dialinecolor}
\node at (7.512500\du,12.378333\du){};
\definecolor{dialinecolor}{rgb}{0.000000, 0.000000, 0.000000}
\pgfsetstrokecolor{dialinecolor}
\node at (5.612500\du,12.325000\du){water};
\definecolor{dialinecolor}{rgb}{0.000000, 0.000000, 0.000000}
\pgfsetstrokecolor{dialinecolor}
\node at (7.425000\du,11.400000\du){leak};
\definecolor{dialinecolor}{rgb}{0.000000, 0.000000, 0.000000}
\pgfsetstrokecolor{dialinecolor}
\node at (9.462500\du,12.375000\du){spray};
\definecolor{dialinecolor}{rgb}{0.000000, 0.000000, 0.000000}
\pgfsetstrokecolor{dialinecolor}
\node at (10.250000\du,11.512500\du){splatter};
\pgfsetlinewidth{0.020000\du}
\pgfsetdash{}{0pt}
\pgfsetdash{}{0pt}
\pgfsetmiterjoin
\definecolor{dialinecolor}{rgb}{0.000000, 0.000000, 0.000000}
\pgfsetstrokecolor{dialinecolor}
\pgfpathellipse{\pgfpoint{10.606250\du}{12.237500\du}}{\pgfpoint{1.793750\du}{0\du}}{\pgfpoint{0\du}{1.225000\du}}
\pgfusepath{stroke}
\definecolor{dialinecolor}{rgb}{0.000000, 0.000000, 0.000000}
\pgfsetstrokecolor{dialinecolor}
\node at (10.606250\du,12.340833\du){};
\pgfsetlinewidth{0.100000\du}
\pgfsetdash{}{0pt}
\pgfsetdash{}{0pt}
\pgfsetmiterjoin
\definecolor{dialinecolor}{rgb}{1.000000, 1.000000, 1.000000}
\pgfsetfillcolor{dialinecolor}
\fill (5.300000\du,10.150000\du)--(5.300000\du,10.225000\du)--(9.750000\du,10.225000\du)--(9.750000\du,10.150000\du)--cycle;
\definecolor{dialinecolor}{rgb}{1.000000, 1.000000, 1.000000}
\pgfsetstrokecolor{dialinecolor}
\draw (5.300000\du,10.150000\du)--(5.300000\du,10.225000\du)--(9.750000\du,10.225000\du)--(9.750000\du,10.150000\du)--cycle;
\pgfsetlinewidth{0.100000\du}
\pgfsetdash{}{0pt}
\pgfsetdash{}{0pt}
\pgfsetmiterjoin
\definecolor{dialinecolor}{rgb}{1.000000, 1.000000, 1.000000}
\pgfsetfillcolor{dialinecolor}
\fill (5.202500\du,14.050000\du)--(5.202500\du,14.125000\du)--(9.652500\du,14.125000\du)--(9.652500\du,14.050000\du)--cycle;
\definecolor{dialinecolor}{rgb}{1.000000, 1.000000, 1.000000}
\pgfsetstrokecolor{dialinecolor}
\draw (5.202500\du,14.050000\du)--(5.202500\du,14.125000\du)--(9.652500\du,14.125000\du)--(9.652500\du,14.050000\du)--cycle;
\pgfsetlinewidth{0.100000\du}
\pgfsetdash{}{0pt}
\pgfsetdash{}{0pt}
\pgfsetmiterjoin
\definecolor{dialinecolor}{rgb}{1.000000, 1.000000, 1.000000}
\pgfsetfillcolor{dialinecolor}
\fill (2.575000\du,10.625000\du)--(2.575000\du,13.500000\du)--(2.625000\du,13.500000\du)--(2.625000\du,10.625000\du)--cycle;
\definecolor{dialinecolor}{rgb}{1.000000, 1.000000, 1.000000}
\pgfsetstrokecolor{dialinecolor}
\draw (2.575000\du,10.625000\du)--(2.575000\du,13.500000\du)--(2.625000\du,13.500000\du)--(2.625000\du,10.625000\du)--cycle;
\pgfsetlinewidth{0.100000\du}
\pgfsetdash{}{0pt}
\pgfsetdash{}{0pt}
\pgfsetmiterjoin
\definecolor{dialinecolor}{rgb}{1.000000, 1.000000, 1.000000}
\pgfsetfillcolor{dialinecolor}
\fill (12.527500\du,10.600000\du)--(12.527500\du,13.475000\du)--(12.577500\du,13.475000\du)--(12.577500\du,10.600000\du)--cycle;
\definecolor{dialinecolor}{rgb}{1.000000, 1.000000, 1.000000}
\pgfsetstrokecolor{dialinecolor}
\draw (12.527500\du,10.600000\du)--(12.527500\du,13.475000\du)--(12.577500\du,13.475000\du)--(12.577500\du,10.600000\du)--cycle;
\pgfsetlinewidth{0.020000\du}
\pgfsetdash{}{0pt}
\pgfsetdash{}{0pt}
\pgfsetbuttcap
{
\definecolor{dialinecolor}{rgb}{1.000000, 0.000000, 0.000000}
\pgfsetfillcolor{dialinecolor}
\pgfsetarrowsstart{latex}
\definecolor{dialinecolor}{rgb}{1.000000, 0.000000, 0.000000}
\pgfsetstrokecolor{dialinecolor}
\pgfpathmoveto{\pgfpoint{6.560588\du}{11.131713\du}}
\pgfpatharc{316}{224}{1.471709\du and 1.471709\du}
\pgfusepath{stroke}
}
\pgfsetlinewidth{0.020000\du}
\pgfsetdash{}{0pt}
\pgfsetdash{}{0pt}
\pgfsetbuttcap
{
\definecolor{dialinecolor}{rgb}{1.000000, 0.000000, 0.000000}
\pgfsetfillcolor{dialinecolor}
\pgfsetarrowsstart{latex}
\definecolor{dialinecolor}{rgb}{1.000000, 0.000000, 0.000000}
\pgfsetstrokecolor{dialinecolor}
\pgfpathmoveto{\pgfpoint{10.606267\du}{11.012515\du}}
\pgfpatharc{311}{224}{1.549530\du and 1.549530\du}
\pgfusepath{stroke}
}
\end{tikzpicture}} }
\setcode{standard}
\vspace{-3em}
\caption{$Syn^2($\RL{mA'}$)$.}
\vspace{-2em}
\label{fig:introsynEx}
\end{center}
\end{figure}

Given a word $w\in L$, 
$Sy(w)=\{u\mid u\in \ss \land\exists s\in \gamma(w)\land~\alpha(u)\cap\alpha(s)\neq\emptyset\}$
is the set of Arabic stems 
directly related to $w$ through the gloss map.
Let $Sy^{i}(w)$ denote stems related to $w$ using the gloss map of order $i$ recursively such that
$Sy^{1}(w) = Sy(w)$ and
$Sy^{i+1}(w)=\{u\mid u\in S\land\exists s\in Sy^{i}(w)\land~\alpha(u)\cap\alpha(s)\neq\emptyset\}$.
Formally, $Syn^k(w) = \bigcup\limits_{i=1}^{k} Sy^{i}(w)$ for $i\in[1 \ldots k]$.
The example in Figure~\ref{fig:introsynEx} illustrates the computation.
Let $w$ denote an input Arabic word \RL{mA'}, which has the gloss \cci{water}, i.e. \cci{water} $\in\alpha(w)$.
$w$ shares this gloss with the stem \RL{n.d.h}, denoted $s_1$, i.e. $s_1\in Sy^{1}(w)$.
Next, the stem \RL{r^s^s}, denoted $s_2$, shares the gloss \cci{spray} with $s_1$, i.e. $s_2\in Sy^{1}(s1)\subset Sy^{2}(w)$.
Therefore, $Syn^2(w)$ relates the words \RL{mA'} and \RL{r^s^s}.

\vspace{-1em}
\subsection{MRE: Morphology-based regular expressions}
\label{subsec:grammar}

Let ${\cal O} = \{ \mathit{isA}, \mathit{contains}\}$ be the set of atomic term 
predicates, where $\mathit{isA}$ and $\mathit{contains}$ denote exact match and containment, respectively.
Also, let ${\cal F} = \{ \pp, \ss, \xx, \PP, \GG, \AC\}$ be the 
set of morphological features where each morphological feature $A\in {\cal F}$ is in 
turn a set of morphological feature values.
Given a word $w$,
a user defined constant feature value $CF\in A$,
and an integer $k, 1\le k\le 7$, 
the following are morphology-based atomic terms (MAT), {\em terms} for short.
\begin{itemize}
  \item $a(w):= \exists m \in M(w).~m=\langle p,s,x,P,G,C\rangle. r \circ CF$
where $\circ \in {\cal O}$, $r \in \{p,s,x,P,G,C\}$, and $r\in A$.
Informally, a solution vector of $w$ exists with
a feature containing or exactly matching the user-chosen feature value $CF$.
\item $a(w) := w \in Syn^k(CF), CF \in \ss$.
  Informally, this checks if $w$ is an extended synonym of a stem $CF$.
  We limit $k$ to a maximum of $7$ since we practically noticed that 
  (1) values above $7$ introduce significant semantic noise and
  (2) the computation is expensive without a bound.
\end{itemize}

A morphology-based Boolean formula (MBF) is of the following form.
\begin{itemize}
  \item $a$ and $\neg a$ are MBF formulae where $a$ is a MAT and $\neg$ is the negation operator. 
  \item $(f \vee g)$ is an MBF where $f$ and $g$ are MBF formulae, 
    and $\vee$ is the disjunction (union) operator. 
\end{itemize}

Moreover, \framework provides $O$ to be a default Boolean formula that tags all {\em other} words in the text that do not match a user defined formula.
We also refer to those words as {\em null} words.

\setcode{utf8}
\setarab
\transfalse
\begin{table}[tb]
\resizebox{\columnwidth}{!} {
	\begin{tabular}{c|c|c|c}
		MBF & description & formula & matches\\ \hline
        N & name of person & $category~=~Name\_of\_Person$ & $n_1,n_2,n_3$ \\ \hline
        P & name of place & $category~=~Name\_of\_Place$ & $p_1,p_2,...,p_7$ \\ \hline
        R & relative position & $stem\in$ \{\RL{قرب},\RL{في},\dots\} & $r_1,r_2,r_3,r_4$ \\ \hline
        U & numerical term & $stem\in$ \{\RL{أول},\RL{ثاني},\dots\} & $u_1,u_2$ \\
    \end{tabular}
}
\vspace{-1em}
\caption{\label{tab:taskMBF}Boolean formulae corresponding to task in Figure~\ref{fig:intromotiv}.}
\vspace{-1em}
\end{table}
\transtrue
\setcode{standard}

Consider the task we discussed in the introduction (Figure~\ref{fig:intromotiv}) 
and recall that we are interested in identifying names of people, names of places, relative positions, and numerical terms.
Table~\ref{tab:taskMBF} presents the defined formulae.
The user denotes the ``name of person'' entities with formula $N$ 
which requires the {\em category} feature in the morphological solution of a word to be {\tt Name\_of\_Person}.
The entities $n_1$, $n_2$, and $n_3$ are matches of the formula $N$ in the text.
Similarly, the user specifies formula $P$ to denote ``name of place'' entities. 
The user specifies formula $R$ to denote ``relative position'' entities, 
and defines it as a disjunction of terms that check for 
solutions matching stems such as \RL{qrb} (``near'') and \RL{fy} (``in'').
Similarly, $U$ denotes numerical terms and is a disjunction of constraints 
requiring the stem feature to belong to a set of stems such as
\RL{'wl}(``first''), \RL{_tAny}(``second''), \dots \RL{`A^sr}(``tenth'').


Next, we define a morphology-based regular expression (MRE) as follows.
\begin{itemize}
\item $m$ is an MRE where $m$ is an MBF.
\item $fg$ is an MRE where $f$ and $g$ are both MRE expressions.
A match of $f$ followed by a match of $g$ satisfies this concatenation operation.
\item $f*,f+,f$\textasciicircum$x,$ and $f?$ are MRE where $f$ is an MRE,
  and are satisfied by zero or more, one or more, up to $x$ matches, 
    and an optional single match of $f$, respectively.
  \item $f\& g,$ (conjunction) and $f|g$ (disjunction) are MRE where 
    $f$ and $g$ are MRE, and are satisfied by the intersection of $f$ and $g$ matches, 
    and the union of the $f$ and $g$ matches, respectively. 
\end{itemize}
We denote by $\ldbrack f\rdbrack$ the set of matches of an MRE $f$.

Back to the example in Figure~\ref{fig:intromotiv}.
We use the formulae defined in Table~\ref{tab:taskMBF} to construct an MRE such as
$(P|N)\!+O?~R~O^\wedge\!2~(P|N|U)+$ where 
$|,+,?,$ and $^\wedge k$ denote disjunction, one or more, zero or one, and
up to $k$ matches, respectively.
The expression specifies a sequence of places or names of persons, 
optionally followed by a null word, 
followed by one relative position, followed by up to two possible null words, 
followed by one or more match of name of place, name of person, or numerical term.
$O?$ and $O^\wedge 2$ are used in the expression to allow for flexible matches.

\setcode{utf8}
\setarab
\transfalse
\begin{figure}[t]
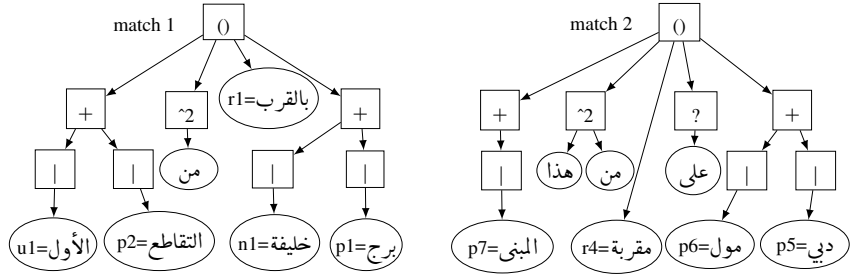

\resizebox{\columnwidth}{!} {
\begin{tabular}{cc}
\relsize{-2} \input{figures/match1.tex} &
\relsize{-2} \input{figures/match2.tex} \\
\end{tabular}
}
\vspace{-1.5em}
  \caption{\label{fig:taskMRE}Matches of regular expression $(P|N)\!+~O?~R~O$\^{}$2~(P|N|U)+$.}
  \vspace{-1em}
\end{figure}
\transtrue
\setcode{standard}

The matching parse trees in Figure~\ref{fig:taskMRE} illustrate two matches of the expression computed by \framework.
The first tree refers to the text 
\RL{brj _hlyfT bAlqrb mn AltqA.t` Al-'wl}(``Khalifa Tower next to the first intersection'').
The second tree refers to the text 
\RL{dby mwl `l_A mqrbT mn h_dA Almbn_A}(``Dubai Mall is located near this building'').
The leaf nodes of the trees are matches to formulae and the internal nodes represent roots to subexpression matches.
For instance, \RL{brj _hlyfT} in match 1 tree corresponds to the subexpression $(P|N)+$.

\vspace{-1em}
\subsection{User-defined relations and actions}
\label{sec:relations-actions}

A relation is defined by the user as a tuple 
$\langle e_1,e_2,r\rangle$ where 
$e_1,e_2,$ and $r$ are identifiers associated with 
subexpressions of an MRE $f$.
Matches of the relation are a set of labeled binary edges
where matches of $e_1$ and $e_2$ are the source and destination nodes
and matches of $r$ are the edge labels.
We denote $\ldbrack \langle e_1,e_2,r\rangle \rdbrack$ to be the set of matches of the corresponding relation, 
and we refer to them as relational entities.

We are interested in constructing the relational entity graph in Figure~\ref{fig:intromotiv}.
Let $e_1$, $o_1$, $r$, $o_2,$, and $e_2$ be identifiers to the subexpressions 
$(P|N)+$, $O?$, $R$, $O\wedge 2$, and $(P|N|U)+$, respectively.
The matches to $e_1$, $r$, $o_2$, and $e_2$ in match 1 (Fig.~\ref{fig:taskMRE}) are \RL{brj _hlyfT} (``Khalifa Tower''), \RL{bAlqrb} (``next''), \RL{mn} (``to''), and \RL{AltqA.t` Al-'wl} (``first intersection'').
Note that there is no match to the optional $O$ formula in match 1.
Similarly, the matches to $e_1$, $o_1$, $r$, $o_2$, and $e_2$ in the second matching tree are \RL{dby mwl} (``Dubai Mall''), \RL{`l_A} (``is located''), \RL{mqrbT} (``near''), \RL{mn h_dA} (``this''), and \RL{Almbn_A} (``building''), respectively.

We define the semantic relations $\langle e_1, e_2, r\rangle$, $\langle r, e_1, o_1\rangle$, and $\langle r,e_2,o_2\rangle$.
Relation $\langle e_1,e_2,r\rangle$ creates the 
edge labeled \cci{next to} between \cci{Khalifa tower} and 
\cci{intersection 1} nodes from match 1, and the 
edge labeled \cci{near} between \cci{Dubai Mall} and \cci{the building} nodes from match 2.
Relation $\langle r,e_1,o_1\rangle$ creates the edge labeled \cci{prep} 
between \cci{Dubai Mall} and \cci{near} nodes from match 2. 
Relation $\langle r,e_2,o_2\rangle$ creates the edge labeled \cci{from} 
between \cci{intersection 1} and \cci{next to} nodes in 
match 1, and the 
edge labeled \cci{from this} between \cci{near} and \cci{the building} 
nodes in match 2.

Moreover, \framework allows advanced users to write C++ code snippets 
to process matches of subexpressions.
Each subexpression can be associated with two computational 
actions: \cci{pre-match} and \cci{on-match}.
\framework provides an API that enriches the actions with detailed access to
all solution features of an expression or a formula match including 
text, position, length, equivalent numerical value when applicable, 
and morphological features.
The API follows a decorator pattern in that it incrementally adds the 
action results to the matching entities.
Once \framework computes all matching parse trees, it traverses each tree to 
execute the user defined \cci{pre-match} actions in pre-order manner
and the \cci{on-match} actions in post-order manner.
This follows an observer pattern that notifies listeners with each
produced match.

\vspace{-1em}
\subsection{\framework simulators}

The set of tag types ${\cal T}$ contains tuples of the form $\langle l,f,d\rangle$ 
where $l$ is a text label with a descriptive name, 
$f$ is an MRE, and $d$ is a visualization legend 
with font and color information.
For the example of Figure~\ref{fig:intromotiv},
$l$ is ``direction'',
$f$ is $(P|N)\!+~O?~R~O^\wedge 2~(P|N|U)+$,
and $d$ is italic.

For each word $t_i\in T, 0\le i < |T|$.
\framework computes a Boolean value for all MBFs.
For example, \RL{brj} matches MBF $P$.
Then, it computes the set of MBF tags
$R_i=\{(t_i,tt)| tt=\langle l,f,d\rangle \wedge
f~\mathit{is~an~MBF} \wedge f(t_i)\} \subseteq T \times {\cal T}$
which tags a word $t_i$ with $\mathit{tt}$ 
iff the MBF $f$ associated with
tag type $\mathit{tt}$ is true for $t_i$. 
The MBF evaluation results in a sequence of tag sets 
$\langle R_0, R_1, \ldots, R_{n-1}\rangle$.
If a word $t_i$ has no tag type match, 
its tag set $R_i$ is by default the singleton $O=\{\mathit{NONE}\}$.
For example, the tag sets for the text in Figure~\ref{f:overview} follows 
$\{ \{NONE\}, \{NONE\}, \{NONE\}, \{NONE\}, \{($\RL{brj}$,P)\}, \{($\RL{xlyfT}$,N)\},\ldots\}$.

\setcode{utf8}
\setarab
\transfalse
\begin{figure}[tb]
\centering
\resizebox{0.75\columnwidth}{!}{
	\relsize{-1} \input{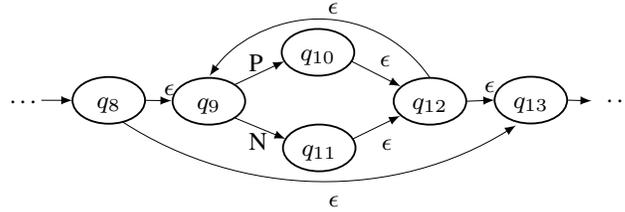}
}
\vspace{-1em}
  \caption{\label{fig:nfaEx}Equivalent NFA of direction expression.}
  \vspace{-1em}
\end{figure}
\transtrue
\setcode{standard}

For each MRE, 
\framework generates its equivalent non-deterministic finite automaton (NFA) in the typical manner~\cite{sipser2012introduction}.
We support the upto operation ($f$\^{}$x$), which is not directly 
supported in~\citename{sipser2012introduction}~\citeyear{sipser2012introduction}, by 
expanding it into a regular expression form; for example 
$f$\^{}$3$ is equivalent to $f?|ff|fff$. 
Consider the example of Figure~\ref{fig:intromotiv} and the
corresponding expression $(P|N)\!+~O?~R~O^\wedge 2~(P|N|U)+$.
Figure~\ref{fig:nfaEx} shows part of the corresponding NFA where
$q_8, q_9, \dots, q_{13}$ represent NFA states,
and edges are transitions based on MBF tags such as 
$P,$ and $N$.
Edges labeled with the empty string $\epsilon$ are non-deterministic.

\framework simulates the generated NFA over the sequence of tag sets matching the MBF formulae.
A simulation match $m$ of an expression $f$ is a parse tree where the root spans the expression, the internal nodes are roots to subexpressions of $f$, and the leaves are matches of the MBF formulae of $f$, e.g. Figure~\ref{fig:taskMRE}.
The sequence of leaf matches forms a vector of tags $\langle r_k,r_{k+1},\dots,r_j\rangle$ 
corresponding to the text sequence 
$\langle t_k,t_{k+1},\dots,t_j\rangle$ where $r_{\ell}\in R_{\ell},0\le k\le \ell \le j < n$.
%
%
If we have more than one match for an expression, 
\framework returns the longest.

Finally, \framework computes the relational entities corresponding to each user defined relation $\ldbrack \langle e_1,e_2,r\rangle \rdbrack$
$\subseteq \ldbrack e_1 \rdbrack \times \ldbrack e_2 \rdbrack \times \ldbrack r \rdbrack$.

\section{\framework GUI}
\label{sec:gui}

\framework~provides a user friendly interface to specify the 
atomic terms, the MBFs, the MREs, 
the tag types, and the legends. 
The GUI also allows the user to modify and correct the 
tag set $R$. 
The GUI allows the user also to compute accuracy results 
that compare different tag sets and that can serve well as 
inter annotation agreement results
when the tag sets come from two human annotators, 
or as evaluation results when comparing with reference tag sets.



\subsection{Tag type Boolean formula editor}
The user writes MBF tag types with the tag type editor introduced in~\citename{JaZaMatar}~\citeyear{JaZaMatar}. 
First the user specifies atomic terms by selecting a feature from ${\cal F}$. 
The user can also choose whether to require an exact match using the
\cci{isA} predicate, or a substring match using the
\cci{contains} predicate option.

The user can add and remove feature values to the atomic terms 
using push buttons. 
A check box in the ``Feature'' column allows negating the term, and the
``Relation'' column switches the predicate between 
\cci{isA} and \cci{contains}. 
The list of feature and value pairs is interpreted as a disjunction to form
the MBF. 
A right pane shows a description of the tag type and a set of legend 
descriptors. 
When the stem or gloss features are selected, the user has the option to 
use the $Syn^k$ feature. 

In the direction extraction task example, the user specifies four MBF-based 
tag types with labels 
$N$, $P$, $R$, and $U$ with  ``name of person'', ``name of place'', 
``relative position'', and ``numerical term'' descriptions, respectively. 
For each MBF, the user selects the morphological features, 
specifies the constant value $CF$, and adds it to the Boolean formula editor. 
%

\subsection{MBF match visualization}

The MBF match visualizer shows color sensitive text view, the tag list view, and the tag description view. 
The tag description view presents the details of the selected tag along with 
the relevant tag type information.
The user can edit the tags using a context sensitive menus. 
\framework GUI also allows manual tag types and corresponding tags 
that are not based on morphological features.
This enables building reference corpora without 
help from the morphological analyzer.

\begin{figure}[tb]
  \centering
  {\setlength{\fboxsep}{0pt}%
  \setlength{\fboxrule}{0.5pt}%
  \fbox{\includegraphics[width=0.995\textwidth]{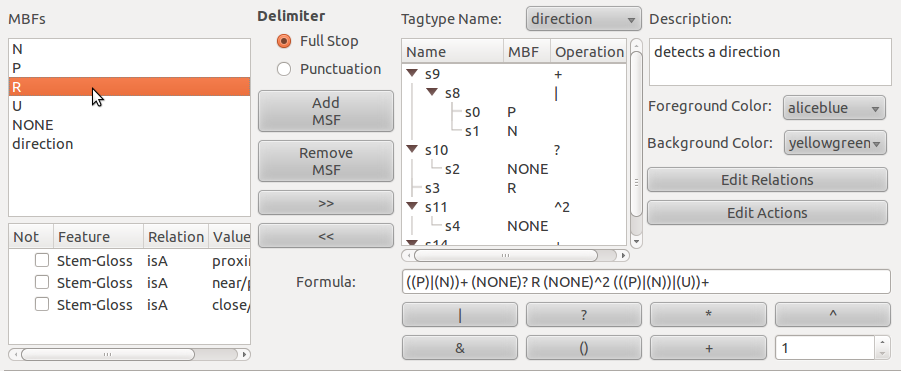}}}
  \vspace{-2em}
  \caption{\label{f:sfe}\framework tag type regular expression editor.}
\end{figure}

\subsection{Tag type regular expression editor}

After interacting with the MBF editor, the user moves to 
specify the regular expressions. 
The MRE editor of Figure~\ref{f:sfe} allows the 
definition of an MRE tag type in a user-friendly manner. 
The user first adds the required MBF formulae 
by selecting a label from ${\cal T}$ under MBFs. 
The Boolean formula of a highlighted tag type is shown in the table on the lower left pane. 
Each selected MBF is associated with an automatic name. 
The user can nest the MRE expression using a tree view of the MRE operations. 
The tree features the name, MBF, and operation for each subexpression. 


To specify a binary operation the user selects two subexpressions and clicks the corresponding
operation button. 
The operations include disjunction, conjunction, zero or one, sequence, zero or more, 
one or more, and up to a user defined constant.
The right pane shows a description of the tag type and a set of legend 
descriptors. 

%

\vspace{-1em}
\subsection{MRE match visualization}

\begin{figure}[tb]
  \centering
  \includegraphics[width=\textwidth]{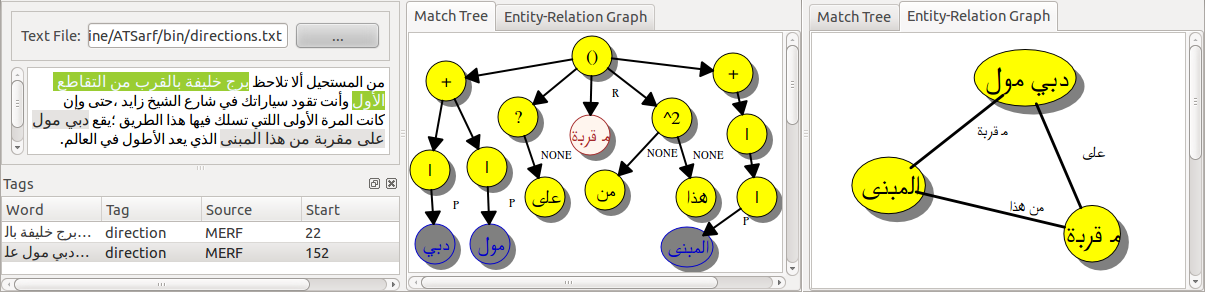}
  \vspace{-2em}
  \caption{\label{fig:treegraph}MRE annotated Text, MRE matching parse tree, and entity-relation graph.}
\end{figure}

While specifying an MRE, the user can interact with the visualization and editor views
to make sure the MRE expresses the intent. 
The color-sensitive text view in Figure~\ref{fig:treegraph} shows 
the highlighted tag matches after the user called the MRE simulator using 
the {\tt Tagtypes} menu. 

The matching parse tree view shows the selected match in a graph view.
Figure~\ref{fig:treegraph} shows the matching parse tree of the direction task 
\RL{dby mwl `l_A mqrbT mn h_dA Almbn_A}(``Dubai Mall is located near this building''). 

\subsection{User defined relation editor}

After the user is satisfied with the MRE matches, 
the user moves to define relations and code actions. 
The relation editor allows the user to define relations 
by specifying $\langle e_1,e_2,r\rangle$ tuples, 
where $e_1$ and $e_2$ denote source and destination entities, and $r$ denotes 
the label.
The editor shows the MRE tree and allows the user to select the subexpressions and select features 
of the matches of the subexpressions to define the three components of the relation. 

A snapshot of the GUI in 
Figure~\ref{fig:treegraph} shows in an interactive graph view
the entity-relation graph of the match of the user defined relation 
extracted from the matching parse tree of the MRE. 
In the computational action editor, an advanced user can 
enter C++ code and use the \framework API to program and process 
subexpression matches. 

\subsection{Analysis}

In the analysis view, the user provides 
two tag sets $R_1$ and $R_2$ and 
two tag type sets ${\cal T}_1$ and ${\cal T}_2$ as input. 
%
The tag type difference view shows the text annotated in three panes: 
(i) the common tag types ${\cal T}_1 \cap {\cal T}_2$,
(ii) the tag types in ${\cal T}_1$ but not in ${\cal T}_2$, 
and (iii) the tag types in ${\cal T}_2$ and not in ${\cal T}_1$.
Similarly, the tag difference view shows $R_1\cap R_2$, $R_1/R_2$ and $R_2/R_1$
in addition to precision, recall and F-measure values. 
The user selects a predicate to compute the metrics from the following predicates:
(1) ``Intersection'': a tag from $R_1$ intersects in text with a tag in $R_2$,
(2) ``Exact'': a tag from $R_1$ exactly matches a tag in $R_2$,
(3) ``A includes B'': a tag from $R_1$ contains a tag from $R_2$, and
(4) ``B includes A'': a tag from $R_2$ contains a tag from $R_1$.

\section{Related Work}
\label{sec:related}

In this section we review the literature on entity and relation IE 
and on automatic and manual annotation techniques 
and compare to \framework.


{\bf Information Extraction.}
The common pattern specification language (CPSL) targets system independent IE specifications~\cite{appelt1998common}. 
\framework extends CPSL with Arabic morphological features, code actions, and user defined relations.
SystemT~\cite{chiticariu2010systemt} aims to overcome the 
limitations of CPSL. 
It is based on an algebraic approach to declarative information extraction, 
uses the declarative annotation query language (AQL), 
and uses an optimizer to generate high performance execution plans for the AQL rules.
\framework supports multiple tags per word, and supports the MRE conjunction operator which overcomes the overlapping annotation problem discussed in SystemT.

TEXTMARKER is a semi-automatic rule-based IE system 
for structured data acquisition~\cite{atzmueller2008rule}.
Both TEXTMARKER and \framework provide the user with GUI editor and result visualizer.

The work in~\citename{urbain2012user}~\citeyear{urbain2012user} presents a user-driven relational model and targets entity and relation extraction. 
The user enters a natural language query, and uses the OpenNLP toolkit to 
extract tags and relations from the query. 
Similar to \framework, the system constructs entities and relations. 

QARAB is an Arabic question answering system that 
takes an Arabic natural language query and provides short answers for it~\cite{hammo2002qarab}. 
QARAB uses traditional information retrieval techniques and an outdated Arabic NLP 
analyzer with limited features of Arabic words compared 
to the morphological analysis of \framework. 

\begin{table}[tb]
\resizebox{\columnwidth}{!}{
\begin{tabular}{l|c|c|c|c|c}
Features & \framework & SystemT & TEXT\linebreak MARKER & Urbain & QARAB\\ \hline
Query type & MRE & AQL & matching \linebreak rules & natural language & natural language \\ \hline
Morphology support& $\checked$ & - & - & OpenNLP & Parser \\ \hline
Relations & $\checked$ & - & - & $\checked$ & - \\ \hline
Actions & $\checked$ & - & - & - & - \\ \hline
Editor & $\checked$ & - & $\checked$ & - & - \\ \hline
Tag visualization & $\checked$ & - & $\checked$ & - & - \\ \hline
Graph visualization & $\checked$ & - & - & - & - \\
\end{tabular}
}
\caption{\label{tab:iecomp}Comparison of \framework with SystemT, 
TEXTMARKER, 
Urbain, 
QARAB.}
\end{table}

Table~\ref{tab:iecomp} summarizes the comparison between \framework and
other systems. \framework differs in that it provides 
code actions, user defined relations, and an interactive 
graph visualization of the relational entities. 
It also differs in that it fully supports Arabic morphological analysis
while only QARAB supports Arabic linguistic features using a parser, and 
the work in~\citename{urbain2012user}~\citeyear{urbain2012user} uses OpenNLP that currently lacks full
support for Arabic morphological features. 
Similar to TEXTMARKER, \framework has the advantage of providing 
a user-friendly interactive interface to edit the entity and relational 
specifications and visualize the results. 

DUALIST is an annotation system for building classifiers for 
text processing tasks using machine learning techniques 
\cite{settles2011closing}. 
\framework doesn't support classification tasks. 
However, \framework provides an interactive GUI where the user can edit MBF and MRE tags.
This interactive environment contributes to the regular expression extraction 
and semantic relation construction which increases the overall accuracy.

Another track in the literature targets specific tasks such as NER using 
statistical and machine-learning techniques such as maximum entropy, 
optimized feature sets and conditional random fields~\cite{benajiba2007anersys,benajiba2008arabic,ekbal2010named,abdelrahman2010integrated}.
Knowledge-based techniques such 
as~\citename{zaghouani2010adapting}~\citeyear{zaghouani2010adapting} and~\citename{traboulsi2009arabic}~\citeyear{traboulsi2009arabic} propose local grammars with morphological stemming. 
\citename{ZaMaHaCicling2012Entity}~\citeyear{ZaMaHaCicling2012Entity} extract entities and events, and relations among them, from Arabic text using a hierarchy of manually built finite state machines driven by morphological features, 
and graph transformation algorithms.
Such techniques require advanced linguistic and programming expertise.

WordNet is a lexical reference system that mimics human lexical memory 
and relates words based on their semantic values and their functional 
categories: nouns, verbs, adjectives, adverbs, and function words~\cite{intro}.
The $Syn^k$ feature in \framework is inspired by WordNet.

{\bf Annotation tools.}
MMAX2 is a manual multi-level linguistic annotation tool with an XML 
based data model~\cite{mmax2}.
BRAT~\cite{brat} and WordFreak~\cite{wordfreak} are manual 
multi-lingual user-friendly web-based annotators that allow the construction 
of entity and relation annotation corpora.
Knowtator~\cite{ogren2006knowtator} is a general purpose incremental text annotation tool 
implemented as a Prot\'eg\'e~\cite{gennari2003evolution} plug-in. 
Prot\'eg\'e is an open-source platform with a suite of tools to construct domain models and knowledge-based applications with ontology. 
However, it doesn't support the Arabic language.

\framework differs from MMAX2, BRAT, WordFreak, and Knowtator in that it is 
an automatic annotator that allows manual corrections 
and sophisticated tag type and relation specifications over
Arabic morphological features.

\citename{kholidy2010towards}~\citeyear{kholidy2010towards} present an overview of annotation tools 
and concludes with a set of rules and guidelines needed in an Arabic annotation alignment tool.
The work in~\citename{dukes2013supervised}~\citeyear{dukes2013supervised} presents a collaborative effort towards morphological and syntactic annotation of the Quran.
\citename{dorr2010interlingual}~\citeyear{dorr2010interlingual} present a framework for interlingual annotation of parallel text corpora with multi-level representations.
\citename{kulick2010consistent}~\citeyear{kulick2010consistent} presents the integration of the Standard Arabic Morphological Analyzer (SAMA) into the workflow of the Arabic Treebank.

The work in~\citename{smrz2004morphotrees}~\citeyear{smrz2004morphotrees} presents 
a customizable general purpose tree editor, with the Arabic MorphoTrees annotations. 
The MorphoTrees present the morphological analyses in a hierarchical organization based on common features. 

Task specific annotation tools such as~\citename{alrahabi2006semantic}~\citeyear{alrahabi2006semantic} use enunciation semantic maps to automatically annotate directly reported Arabic and French speech.
AraTation is another task specific tool for semantic annotation of 
Arabic news using web ontology based semantic maps~\cite{saleh2009aratation}.
We differ in that \framework is general, and not task specific, and it uses 
morphology-based features as atomic terms.
Fassieh is a commercial Arabic text annotation tool that enables the production of large Arabic text corpora~\cite{attia2009fassieh}.
The tool supports Arabic text factorization including morphological analysis, POS tagging, 
full phonetic transcription, and lexical semantics analysis in an automatic mode.
Fassieh is not directly accessible to the research community and requires commercial licensing.
\framework is open source and differs in that it allows the user to build tag types and extract entities and relations from text.

\vspace{-1em}
\section{Results}
\label{sec:results}

In this section we evaluate \framework with four case studies. 
We perform a survey-like evaluation where developers manually 
built task specific information extraction tools for the case studies 
and other developers built equivalent \framework tools. 
The aim of the comparison is to showcase that \framework enables 
fast development of linguistic applications with similar accuracy 
and a reasonable affordable overhead in computational time. 
We report development time,
size of developed code versus size of grammar,
running time, and 
precision-recall as metrics
of cost, complexity, overhead, and accuracy,
respectively. 

We survey three case studies from the literature: 
(1) narrator chain, (2) temporal entity, and (3) genealogy entity extraction 
tasks, and we use the reported development time for the task specific 
techniques proposed in
ANGE~\cite{ZaMaFlairs2012HadithBio}, 
ATEEMA~\cite{ZaMa2012IJCLATime},  and
GENTREE~\cite{ZaMaHaCicling2012Entity}, respectively. 
We also compare a \framework number normalization task to 
a task specific implementation. 

We evaluated ANGE with \textit{Musnad Ahmad}, a hadith book, where we constructed an annotated golden reference containing 1,865 words.
We evaluated ATEEMA with articles from issues of  the Lebanese \textit{Al-Akhbar} newspaper where we constructed an annotated golden reference containing 1,677 words. 
For the genealogical tree extraction we used an extract from the Genesis biblical text with 1,227 words.
Finally, we used an annotated article from the Lebanese \textit{Assafir} newspaper with 1,399 words to evaluate the NUMNORM case study\footnote{Available at~\url{http://www.assafir.com} and 
\url{http://www.al-akhbar.com}.}. 
In the online appendix\footnote{Available at~\url{http://research-fadi.aub.edu.lb/pdfs/merfappendix.pdf}.}, we report on eight additional \framework case studies.
Manual annotators inspected the outcome and provided corrections where tools 
made mistakes.
The corrections form the manual gold annotation that we compared against.

\begin{table}[tb]
  \centering
  \resizebox{\columnwidth}{!}{
    \begin{tabular}{l|l|l|l|ll|l}
     \toprule
      \multirow{2}{*}{Task} & Size & Development& Run & \multicolumn{2}{c|}{Accuracy} & \multirow{2}{*}{Ease of Composition}\\
      & (words) & time & time(s) & Recall & \multicolumn{1}{c|}{Precision} & \\
    \midrule
      ANGE & 1,865  
    & 2 months & 1.79 & 0.99 & 0.99 & 3,000+ lines of code\\
      \framework &  & 3 hours & 7.24 & 0.99 & 0.93  & 8 MBFs and 4 MREs\\
    \midrule
      ATEEMA & 1,677 & 1.5 months & 2.53 & 0.88  & 0.89  & 1,000+ lines of code  \\
      \framework & & 3 hours & 3.14 & 0.91  & 0.81  & 3 MBFs and 2 MREs\\
    \midrule
      Genealogy tree & 1,227 
    & 3 weeks & 0.74 & 0.96 & 0.98 & 3,000+ lines of code\\
      \framework &  & 4 hours & 2.28 & 0.84 & 0.93  & 3 MBFs and 3 MREs\\
    \midrule
      NUMNORM &  1,399 & 1 week & 0.32 & 0.91 & 0.93  & 500 lines of code\\
      \framework &  & 1 hour & 1.53 & 0.91 & 0.90 & 3 MBFs/1 MRE/57 lines\\
    \bottomrule
    \end{tabular}%
    }
   \caption{\label{tab:results}\framework compared to task specific applications.}
\end{table}%

Table~\ref{tab:results} reports the development time,
extraction runtime, recall and precision 
of the output MRE tags, 
the size of the task in lines of code or in number of \framework rules, 
for both the standalone task specific and the \framework implementations.
The development time measures the time required for 
developing the case study.
For instance, ANGE~\cite{ZaMaFlairs2012HadithBio} required two 
months of development by a research assistant with 6 and 14 hours of 
course work and teaching duties, respectively.
Recall refers to the fraction of the entities correctly detected against the
total number of entities. 
Precision refers to the fraction of correctly
detected entities against the total number of extracted entities. 

Table~\ref{tab:results} provides runtime results of \framework 
compared to the task specific implementations while running 
MBF and MRE simulations jointly.
This is a rough estimate of the complexity of the \framework simulator. 
The complexity of the MBF simulation is the total number of morphological 
solutions for all the words multiplied by the number of user-defined MBFs.
We do not provide a limit on the number of user defined formulae.
In practice, we did not encounter more than ten formulae per case study.
As for the complexity of MRE simulation, converting the rules into 
non-deterministic finite state machines (NDFSM) is done once. 
Simulating an NDFSM over the MBF tags is potentially exponential. 
In practice, all our case studies terminated within a predetermined 
time bound of less than 30 minutes. 
\framework required reasonably more runtime than the task specific 
implementations and reported acceptable and 
slightly less precision metrics with around
the same recall.



Table~\ref{tab:results} shows that \framework has a clear advantage over 
task specific techniques in the effort required to develop the application at 
a reasonable cost in terms of accuracy and run time. 
Developers needed three hours, three hours, four hours, and one hour 
to develop the narrator chain, temporal entity, genealogy, and number 
normalization case studies using \framework, respectively. 
However, the developers of ANGE, ATEEMA, GENTREE, and 
NUMNORM needed two months, one and a half months, 
three weeks, and one week, respectively. 
\framework needed eight MBFs and four MREs for narrator chain, 
three MBFs and two MREs for temporal entity, three MBFs and three MREs for 
genealogy, and three MBFs, one MRE, and 57 lines of code actions for the number normalization tasks. 
However, ANGE, ATEEMA, GENTREE, and NUMNORM required 
3,000+, 1,000+, 3,000+, and 500 lines of code, respectively.


\setcode{utf8}
\setarab
\begin{table}[tb!]
  \centering
  \begin{Verbatim}[xleftmargin=1.5cm,fontsize=\relsize{-1},commandchars=\\\{\},codes={\catcode`$=3 \catcode`_=8}]
name:   PN ((MEAN)? PN)*;
nar:    name ((NONE)^3 FAM (NONE)^3 name)*;
pbuh:   BLESS GOD UPONHIM GREET;
nchain: ($s_1=$TOLD $s_2=$nar)+ ((PN|FAM|NONE)^8 pbuh)?
\end{Verbatim}
\resizebox{0.8\columnwidth}{!}{
    \begin{tabular}{|c|c|c|c|c|c|c|c|c|c|}
    \toprule 
    \notrrl{القعقاع} & \notrrl{بن} & \notrrl{عمارة} & \notrrl{عن} & \notrrl{جرير} & \notrrl{حدثنا} & \notrrl{سعيد} & \notrrl{بن} & \notrrl{قتيبة} & \notrrl{حدثنا} \\
    \midrule 
    \noarrl{القعقاع} & \noarrl{بن} & \noarrl{عمارة} & \noarrl{عن} & \noarrl{جرير} & \noarrl{حدثنا} & \noarrl{سعيد} & \noarrl{بن} & \noarrl{قتيبة} & \noarrl{حدثنا} \\
    \midrule
    PN    & FAM   & PN    & TOLD  & PN    & TOLD  & PN    & FAM   & PN    & TOLD \\
    \midrule
    name & & name & & name & & name &  & name & \\
    \midrule
    \multicolumn{3}{|c|}{nar} &       & nar   &       & \multicolumn{3}{c|}{nar} &  \\
    \midrule
    \multicolumn{10}{|c|}{nchain}
    \\
    \bottomrule
    \end{tabular}%
}
\caption{\label{tab:nchain}Narrator chain example.}
 \vspace{-1em}
\end{table}
\setcode{standard}

\subsection{Narrator chain case study}
A narrator chain is a sequence of narrators referencing each other. 
The chain includes proper nouns, paternal entities, and referencing entities. 
ANGE uses Arabic morphological analysis, finite state machines, and graph transformations 
to extract entities and relations including narrator chains~\cite{ZaMaFlairs2012HadithBio}.

\transfalse
Table~\ref{tab:nchain} presents the MREs for the narrator chain case study. 
MBF \cci{PN} checks the abstract category {\tt Name of Person}. 
MBF \cci{FAM} denotes ``family connector'' and checks the stem gloss ``son''. 
MBF \cci{TOLD} denotes referencing between narrators and checks the disjunction of 
the stems \RL{.hd_t}(``spoke to''), \RL{`n}(``about''), \RL{sm`}(``heard''), \RL{'_hbr}(``told''), and \RL{'nb-'}(``inform''). 
MBF \cci{MEAN} checks the stem \RL{`ny}(``mean''). 
MBFs \cci{BLESS}, \cci{GOD}, \cci{UPONHIM}, and \cci{GREET} check the 
stems \RL{.sll_A}, \RL{Al-ll_ah}, \RL{`ly}, and \RL{sllm}, respectively. 
\transtrue

MRE {\em name} is one or more \cci{PN} tags optionally followed 
with a \cci{MEAN} tag. 
MRE \cci{nar} denotes narrator which is a complex Arabic name
composed as a sequence of Arabic names (\cci{name}) 
connected with family indicators (\cci{FAM}). 
The \cci{NONE} tags in \cci{nar} allow for unexpected words 
that can occur between names. 
MRE \cci{pbuh} denotes a praise phrase often associated with 
the end of a hadith (``peace be upon him''), 
and is the satisfied by the sequence of
\cci{BLESS}, \cci{GOD}, \cci{UPONHIM}, and \cci{GREET} tags. 
MRE \cci{nchain} denotes narrator chain, 
and is a sequence of narrators (\cci{nar})
separated with \cci{TOLD} tags, and optionally followed
by a \cci{pbuh} tag. 

The first row in Table~\ref{tab:nchain} is an example narrator chain,
the second is the transliteration, the third 
shows the MBF tags. Rows 4, 5, and 6 show the 
matches for \cci{name}, \cci{nar}, and \cci{nchain},
respectively.
\framework assigns the symbols $s_1$ and $s_2$ for the 
MRE subexpressions \cci{TOLD} and \cci{nar}, respectively. 
We define the relation $\langle s_2,s_2',s_1\rangle$ 
to relate sequences of narrators with edges labeled by the tags of \cci{TOLD} where 
$s_2'$ denotes the next match of \cci{nar} in the one or more MRE subexpression.
%
Table~\ref{tab:mbfer} shows that \framework detected almost all the MBF matches 
with 99\% recall and 85\% precision and 
extracted user-defined relations with 98\% recall and 99\% precision.


\begin{table}[tb]
  \centering
  \resizebox{0.8\columnwidth}{!}{
    \begin{tabular}{l|c|c|c|c}
     \toprule
     \multirow{2}{*}{Task} & \multicolumn{2}{c|}{MBF accuracy} & \multicolumn{2}{c}{relation accuracy}\\
     & Recall & Precision & Recall & Precision \\
    \midrule
    Narrator chain & 0.99 & 0.85 & 0.99 & 0.98 \\
    Number normalization & 0.99 & 0.99 & 0.97 & 0.95 \\
    Temporal entity & 0.99 & 0.52 & 0.98 & 0.89 \\
    Genealogy tree & 0.99 & 0.75 & 0.81 & 0.96 \\
    \bottomrule
    \end{tabular}%
    }
  \caption{\label{tab:mbfer}\framework MBF and user-defined relation accuracy.}
 \vspace{-1.5em}
\end{table}%

\vspace{-1.5em}
\subsection{Temporal entity extraction}
\vspace{-1em}

Temporal entities are text chunks that express temporal information.
Some represent absolute time such as~\RL{Al_hAms mn 'Ab 2010}$\!\!$.
Others represent relative time such as~\RL{b`d _hmsT 'ayAm}$\!\!\!$, and quantities such as~\RL{14 ywmA}$\!$.
{\em ATEEMA} presents a temporal entity detection technique for the Arabic language using 
morphological analysis and finite state transducers~\cite{ZaMa2012IJCLATime}. 
Table~\ref{tab:mbfer} shows that \framework detected almost all the MBF matches with 99\% recall, 
however it shows low precision (52\%). 
As for the semantic relation construction, \framework presents a 98\% recall and 89\% precision.

\vspace{-1.5em}
\subsection{Genealogy tree}
\vspace{-1em}

Biblical genealogical lists trace key biblical figures such as Israelite kings and
prophets with family relations. 
The family relations include wife and parenthood. 
A sample genealogical chunk of text is \RL{w wld hArAn lw.tA} 
meaning ``and Haran became the father of Lot''.
GENTREE~\cite{ZaMaHaCicling2012Entity} 
automatically extracts the genealogical family trees using morphology, 
finite state machines, and graph transformations. 
Table~\ref{tab:mbfer} shows that \framework detected 
MBF matches with 99\% recall, and 75\% precision, and
extracted relations with 81\% recall and 96\% precision.

\vspace{-1em}
\subsection{Number normalization}
\label{subsec:numnorm}

\newcommand*{\fvtextcolor}[2]{\textcolor{#1}{#2}}
\begin{figure}[tb]
\centering
  \begin{tabular}{p{5.5cm}p{5.8cm}}
\begin{Verbatim}[fontsize=\relsize{-1},frame=single,label=TMB algorithm,commandchars=\\\[\]] 
cout << \fvtextcolor[red][$s1.text];
if(isHundred) {
  if(current != 0) {
    previous += current;
  }
  current = currentH * \fvtextcolor[red][$s1.number];
  currentH = 0;
  isHundred = false;
  isKey = true;
} else if(current == 0) {
  current = \fvtextcolor[red][$s1.number];
  isKey = true;
} else if(!isKey) {
  isKey = true;
  current = current * \fvtextcolor[red][$s1.number];
} else {
  previous += current;
  current = \fvtextcolor[red][$s1.number];}
\end{Verbatim}
&
\begin{Verbatim}[fontsize=\relsize{-1},frame=single,label=DT algorithm,commandchars=\\\[\]] 
if(isHundred) {currentH += \fvtextcolor[red][$s0.number];
} else if(current == 0) {
  current = \fvtextcolor[red][$s0.number];
} else if(isKey) {
  previous += current;
  current = \fvtextcolor[red][$s0.number];
} else {current += \fvtextcolor[red][$s0.number]; }
isKey = false;
\end{Verbatim}
\begin{Verbatim}[fontsize=\relsize{-1},frame=single,label=H algorithm,commandchars=\\\[\]] 
isHundred = true;
if(current == 0)  {
  currentH = \fvtextcolor[red][$s2.number];
} else if(!isKey) {
  currentH = current * \fvtextcolor[red][$s2.number];
  current = 0;
} else {currentH = \fvtextcolor[red][$s2.number];}
isKey = false;
\end{Verbatim}
\\ 
\end{tabular}
 \vspace{-2em}
\caption{Actions for TMB, DT, and H MRE expressions.}
\vspace{-1.5em}
\label{fig:numnormalgo}
\end{figure}

We implemented a number normalization extractor using \framework and 
compared it with {\em NUMNORM}, a 
C++ implementation for number normalization. 
First, we defined the MBFs \cci{DT}, \cci{H}, and \cci{TMB}
to denote (1) digits and tens, (2) hundreds, and (3) 
thousands, millions, and billions, respectively.
The \cci{num} MRE 
\cci{(DT|TMB|H)+} is one or more \cci{DT}, \cci{TMB}, or \cci{H} tags. 
\framework assigns the symbols $s_1$, $s_2$, and $s_3$ 
for the subexpressions \cci{DT}, \cci{TMB}, and \cci{H}, respectively. 
Figure~\ref{fig:numnormalgo} shows the actions associated with the \cci{DT}, \cci{TMB}, and \cci{H} subexpressions that cumulatively compute the numeric value of the numeric expression match.
The actions use \framework API to access features of the matches such 
as text (\cci{\$s1.text}) and numeric 
value (\cci{\$s1.number}) of literal numbers such as numbers from one to ten.
Table~\ref{tab:mbfer} shows high accuracy in MBF tagging and relation extraction 
with 99\% and 97\% recall and 99\% and 95\% precision, respectively. 

\vspace{-1.5em}
\subsection{Discussion}
\label{subsec:discuss}
\label{sec:discuss}

The results show that \framework 
provides a friendly environment to develop entity and relational
entity extraction tasks with acceptable 
accuracy and runtime overheads compared to task specific applications. 
\framework requires the user to understand and interact with 
basic linguistic concepts such as readable values of morphological 
features, sequences, repetitions, and bounded repetitions. 
The user interacts with the MBF editor to specify basic concepts
and visualize their matches over highlighted text. 
Then, the user interacts with the MRE editor to specify 
sequences of the concepts and visualize the matches
in a graph, in conjunction with the highlighted text.

The two levels of interaction allow the user to separate between concepts 
that relate to word features, and more sophisticated entities 
that relate to sequences and context. 
The MBF, MRE, and user defined relations 
can be used to generate large annotated corpora in a fast manner. 
\framework visualization can be used later to refine the annotation.
%
The case studies showed that 
\framework requires some linguistic expertise to successfully execute the 
tasks.
In contrast, the case specific implementations require more sophisticated 
linguistic and programming expertise to attain similar results.

We notice that ANGE, ATEEMA, and Genealogy tree report higher precision than \framework. 
This is mainly due to their capacity 
to learn words and relations that may not have a match in the 
morphological analyzer based on co-occurrence relations. 
For example, the sequence $p_1 t_1 p_2$ where $p_1$ and 
$p_2$ are persons and $t_1$ is a tell relationship helps
indicate that $x$ is a tell relationship in $p_1 x p_2$ 
even if the morphological analyzer did not return the required
feature for $x$ to match a tell relationship. 
\framework does not have that capacity yet unless it is
encoded in the C++ actions.


\vspace{-1em}
\section{Conclusion}
\label{sec:conclusion}

In this work, we present a morphology-based entity and relational entity extraction framework for Arabic text.
\framework provides a friendly interface where the user defines tag types 
and associates them with regular expressions defined over Boolean formulae.
The Boolean formulae are in turn defined over matches of Arabic morphological features and
a novel extended synonymy feature ($Syn^k$).
\framework allows the user to associate code actions with each regular subexpression and to define semantic relations between subexpressions. 
We evaluate \framework with several case studies and compare with existing application-specific 
techniques.
The results show that \framework requires shorter development time and effort compared 
to existing techniques and produces reasonably accurate results within a reasonable 
overhead in run time. 
In the future, \framework will support user-defined cross-reference predicates, 
and will infer morphological features from relevant example words to express a concept.


\vspace{-1.5em}
\section{Acknowledgment}
\vspace{-1em}
The authors would like to thank the Lebanese National Council for Scientific Research (CNRS) for their support.

\vspace{-1em}
\bibliographystyle{biblio-hermes}
{\bibliography{merf}}
\end{document}